\documentclass[letter,aps,reprint,superscriptaddress]{revtex4-1}
\usepackage[english]{babel}
\usepackage{amsmath}
\usepackage{amssymb}
\usepackage{hyperref}
\usepackage{braket}
\usepackage{graphicx}
\usepackage{textgreek}
\usepackage{siunitx}
\usepackage{xcolor}
\usepackage{placeins}
\usepackage[normalem]{ulem}

\usepackage{comment}

\newcommand{\ad}{\hat{a}^{\dag}}
\renewcommand{\a}{\hat{a}}

\newcommand{\Ec}{E_0}
\newcommand{\kp}{\mathbf{k}_p}
\newcommand{\kc}{\mathbf{k}_c}
\renewcommand{\r}{\mathbf{r}}

\renewcommand{\k}{\mathbf{k}}
\newcommand{\Vp}{V_\mathrm{p}}

\newcommand{\ez}{\mathbf{e}_z}

\newcommand{\beginsupplement}{%
	\setcounter{table}{0}
	\renewcommand{\thetable}{S\arabic{table}}%
	\setcounter{figure}{0}
	\renewcommand{\thefigure}{S\arabic{figure}}%
      \setcounter{equation}{0}
	\renewcommand{\theequation}{S\arabic{equation}}%
}

\newcommand{\ETH}{Institute for Quantum Electronics, Eidgen\"ossische Technische Hochschule Z\"urich, Otto-Stern-Weg 1, 8093 Zurich, Switzerland}

\begin{document}

\title{Measuring the dynamics of a first order structural phase transition between two configurations of a superradiant crystal
}
\date{\today}

\author{Xiangliang Li}
\affiliation{\ETH}
\author{Davide Dreon}
\affiliation{\ETH}
\author{Philip Zupancic}
\affiliation{\ETH}
\author{Alexander Baumg\"artner}
\affiliation{\ETH}
\author{Andrea Morales}
\affiliation{\ETH}
\author{Wei Zheng} \affiliation{T.C.M.~Group, Cavendish Laboratory, University of Cambridge, J.J.~Thomson Avenue, Cambridge CB3 0HE, United Kingdom}
\affiliation{Hefei National Laboratory for Physical Sciences at the Microscale and Department of Modern Physics, University of Science and Technology of China, Hefei 230026, China}
\affiliation{CAS Center for Excellence in Quantum Information and Quantum Physics,  University of Science and Technology of China, Hefei 230026, China}
\author{Nigel R.~Cooper} \affiliation{T.C.M.~Group, Cavendish Laboratory, University of Cambridge, J.J.~Thomson Avenue, Cambridge CB3 0HE, United Kingdom}
\author{Tobias Donner}
\email{donner@phys.ethz.ch}\affiliation{\ETH}
\author{Tilman Esslinger}
\affiliation{\ETH}

\begin{abstract}
We observe a structural phase transition between two configurations of a superradiant crystal by coupling a Bose-Einstein condensate to an optical cavity and applying imbalanced transverse pump fields. We find that this first order phase transition is accompanied by transient dynamics of the order parameter which we measure in real-time. The phase transition and the excitation spectrum can be derived from a microscopic Hamiltonian in quantitative agreement with our experimental data.
\end{abstract}

\maketitle
Structural phase transitions between different crystal configurations play an important role in the description of materials. They arise from a delicate balance of competing internal forces and can be complex to describe owing to their intrinsically nonlinear character. The study of the transition dynamics is especially challenging, due to the very short time scales determining the process in solid state systems \cite{Beaud:2014aa, johnson2012femtosecond, eichberger2010snapshots, Yusupov:2010aa}. Beyond condensed matter systems, quantum structural phase transitions have also been studied in ion crystals \cite{Walther1992, Raizen1992, Shimshoni2011} at effectively zero temperature.

\begin{figure}[t!]
\centering
\includegraphics[width=0.96\columnwidth]{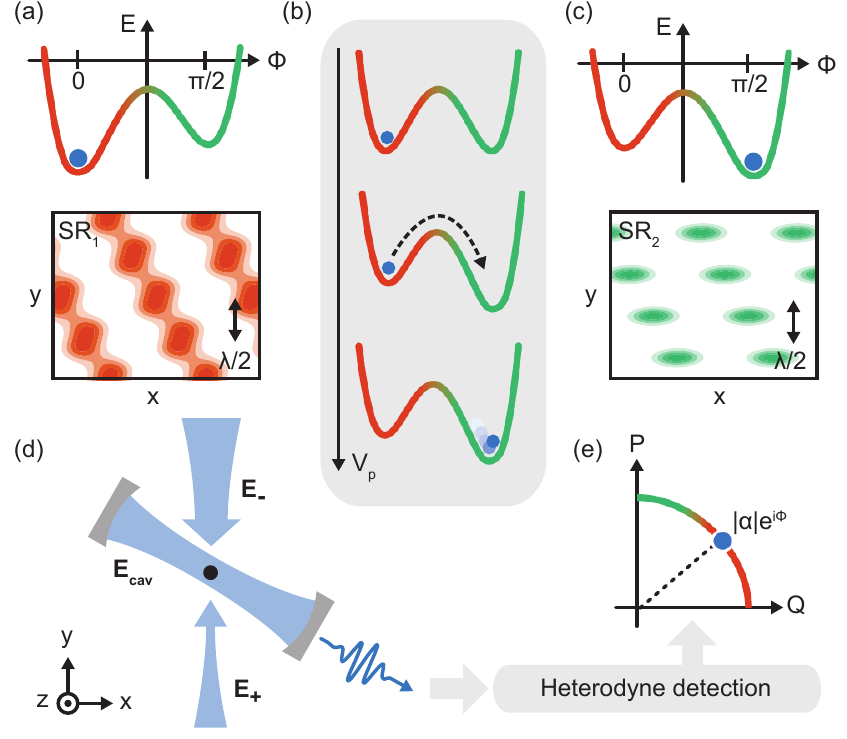}
\caption{Conceptual model and schematics of the experiment.  (a, c) The free energy $E$ as a function of the phase $\phi$ of the intra-cavity light field supports two distinct minima. They correspond to two crystal structures SR$_1$ and SR$_2$ with different symmetries as sketched in real space in red and green. (b) As the free energy deforms during the experimental sequence, swapping the local and the global minima, the system undergoes a first order phase transition. The excess energy from the metastable state results in a damped oscillation of the order parameter around the new global minimum. (d) Two imbalanced counter-propagating pump beams $\mathbf{E}_\pm$ couple the Bose-Einstein condensate to the quadratures $Q$ and $P$ of the intra-cavity electric field $\mathbf {E}_c$. Each quadrature corresponds to a different interference pattern of the electric fields. (e) Using a heterodyne detector we measure amplitude $|\alpha|$ and phase $\phi$ of the light field leaking from the cavity. From the phase we reconstruct to which quadrature the atoms are coupled, and hence which crystal structure they acquire.
}
\label{fig1}
\end{figure}

In quantum simulations with ultracold atoms loaded into optical lattices \cite{Lewenstein2012, Bloch2008Many}, the lattice structure is dictated by the externally applied laser fields, which is strength but also limitation of this approach.  For example, the crystallization process itself, or a structural phase transition between different crystal configurations, cannot be studied. Such phenomena can however be addressed with dynamical lattice potentials that emerge inside optical cavities coupled to driven atoms \cite{Ritsch2013}.

In such many-body cavity QED settings, atoms are placed into an initially unoccupied cavity mode and illuminated by an external pump laser field. Above a critical pump strength, the atoms lower the total energy of the system by crystallizing into a superradiant pattern that supports Bragg scattering of photons into the cavity mode. The interference between the classical pump laser field and the emerging self-consistent quantized cavity field then gives rise to a dynamic potential that enforces the atomic pattern formation. This approach has been used to study diverse aspects of crystallization phase transitions between an initially unordered phase and a superradiant crystal \cite{Black:2003aa, Baumann:2010aa, Ritsch2013, Schmidt2014Dynamical, Klinder:2015aa, Leonard2017Supersolid, KollAr:2017aa}. Yet, first-order structural phase transitions between two superradiant configurations of distinct geometry have not been observed.

The challenge in implementing such transitions in many-body cavity QED simulations lies in the fact that multiple cavity modes need to be involved \cite{Gopalakrishnan2009Emergent}. In this Letter, we demonstrate that a structural phase transition can also be induced in an experimentally simpler way by coupling to both quadratures of a single cavity mode. Monitoring the light field leaking out of the cavity, we observe the associated rapid jump, the oscillation, and the relaxation of the order parameter.

In our experiment, we induce a first order  transition between two different emergent crystalline configurations that arise in the atomic density of a Bose Einstein condensate (BEC) coupled to a high-finesse Fabry-P\'erot optical cavity. The two crystal structures have different symmetries of the wave function, and correspond to distinct minima of the free energy of the system (Fig.~\ref{fig1}). The microscopic origin of the two structures can be understood from the interaction between the atoms and the light, which is described by the Hamiltonian $H_{A-L}  = - \alpha_s  \,\mathbf {E}^*\cdot \mathbf {E}$, where $\alpha_s$ is the scalar atomic polarizability and $ \mathbf {E}$ is the total electric field.
The BEC is placed at the mode centre of a high-finesse optical cavity and exposed to an off-axis pump laser beam, see Fig.~\ref{fig1} (d). The total electric field is the sum of the cavity and the pump fields  $\mathbf {E} = \mathbf {E}_c +  \mathbf{E}_p$. The cavity is initially in the vacuum state but can be populated by Raman processes where photons are scattered via the atoms from the pump into the cavity (and vice versa). In order to scatter light constructively, the atoms have to organize in a periodic structure that obeys the Bragg condition and comes at a cost of kinetic energy. This is thus only possible above a critical power of the pump beam, where the overall energy is lowered by atomic self-organization~\cite{Domokos:2002aa,Ritsch2013}, and the system becomes superradiant. At the phase transition, the BEC spontaneously breaks a discrete translational symmetry~\cite{Baumann2011}.
In contrast with previous self-organization studies, that relied either on a standing wave~\cite{Black:2003aa,Baumann:2010aa,Klinder:2015aa,KollAr:2017aa} or a running wave~\cite{Arnold:2012aa,Bux:2011aa,Kesler:2014aa} pump beam, in this Letter we employ two unbalanced counter-propagating beams \cite{Mivehvar2019Cavity}, that is $\mathbf{E}_p = \mathbf{E}_+ +\mathbf{E}_-$, with $\mathbf{E}_\pm= E_\pm e^{\pm i\k_p\r}\ez$. Here, $E_{\pm}$ are the electric field amplitudes, $\ez$ is the polarization vector, $\k_p$ is the wave vector with $|\mathbf{k}_{p}|=2\pi/\lambda$ and $\lambda$ is the wavelength of the light. The beam in the $-\k_p$ direction is the retro-reflected $+\k_p$ beam, whose focus position allows us to tune the imbalance parameter $\gamma = (E_+ /E_-)^{1/2}$.
The two beams interfere, creating a standing wave with an offset. The standing wave corresponds to an optical lattice depth $V_p = -\alpha_s E_+ E_-$, which we use as control parameter for the phase transition. The interference between the cavity and the pump electric fields generates two potential energy terms with different symmetry, that give rise to the two different patterns for the atomic density. This follows from writing the interaction Hamiltonian $\hat H_{A-L}$ as~\cite{SM}:
\begin{equation}
\begin{split}
\hat H_{A-L} (\r)&= V_p \cos^2(\k_p\r) +\hat  V_c \cos^2(\k_c\r)  \\
& + \hat V_1 \cos(\k_p\r)\cos(\k_c\r) +\hat V_2 \sin(\k_p\r)\cos(\k_c\r).
 \end{split}
 \label{eq:HamiltonianAR}
 \end{equation}
Here, $\mathbf{k}_{c}$ is the wave vector along the cavity axis, with $|\mathbf{k}_{c}|=|\mathbf{k}_{p}|$. $\hat V_c  = \hbar U_0 \ad\a$ is the potential of the quantized cavity lattice and
\begin{equation}
\begin{split}
\hat V_1 & = \left(\gamma+\frac{1}{\gamma}\right) \left(\hbar U_0 V_p \right)^{1/2} (\ad+\a), \\
\hat V_2 & =- i \left(\gamma-\frac{1}{\gamma}\right)\left(\hbar U_0 V_p \right)^{1/2} (\ad-\a)
 \end{split}
 \label{eq:V}
\end{equation}
are the two possible interference terms between pump and cavity, where $\ad$ ($\a$) is the creation (annihilation) operator for a cavity photon, and $U_0$ is the AC Stark shift from a single cavity photon, or equivalently, the single-atom dispersive shift. While both quadratures, $\hat{V}_{1,2}$, couple from $\mathbf{k}= \mathbf{0}$ to $\mathbf{k}_- = \mathbf{k}_{p} - \mathbf{k}_{c}$, the quadrature $\hat{V}_1$ couples more strongly to the second ($p$) band than to the first ($s$) band (and vice versa). The different Bloch wavefunctions of the $s$- and $p$-bands mean that the SR phases in these two quadratures will have different spatial structures. The two possible configurations for the lattices correspond to the two minima of the free energy in Fig.~\ref{fig1}, and are either $\braket{\hat{V}_{1}}\neq 0$ and $\braket{\hat{V}_2}= 0$ or  $ \braket{\hat{V}_{2}}\neq 0$ and $\braket{\hat{V}_1}= 0$. They will be referred to as superradiant phases SR$_1$ and SR$_2$, respectively.
All potentials except $V_p$ have a nonzero value only in the superradiant phases, i.e. when $\braket{\hat a} \neq 0$. In addition, $\hat{V}_1$ and $\hat{V}_2$ are coupled to the orthogonal quadratures $Q = \frac{1}{\sqrt{2}}\braket{\ad+\a}$ and $P = \frac{i}{\sqrt{2}}\braket{\ad-\a}$ of the cavity field. Therefore, using the complex-valued expectation value of the intra-cavity field $\braket{\hat a}= \alpha = |\alpha| e^{i\phi}$ as an order parameter, it is possible to observe not only the transition to a superradiant phase, but also to distinguish SR$_1$ from SR$_2$ via the phase of the light field.

\begin{figure}[b!]
\centering
\includegraphics[]{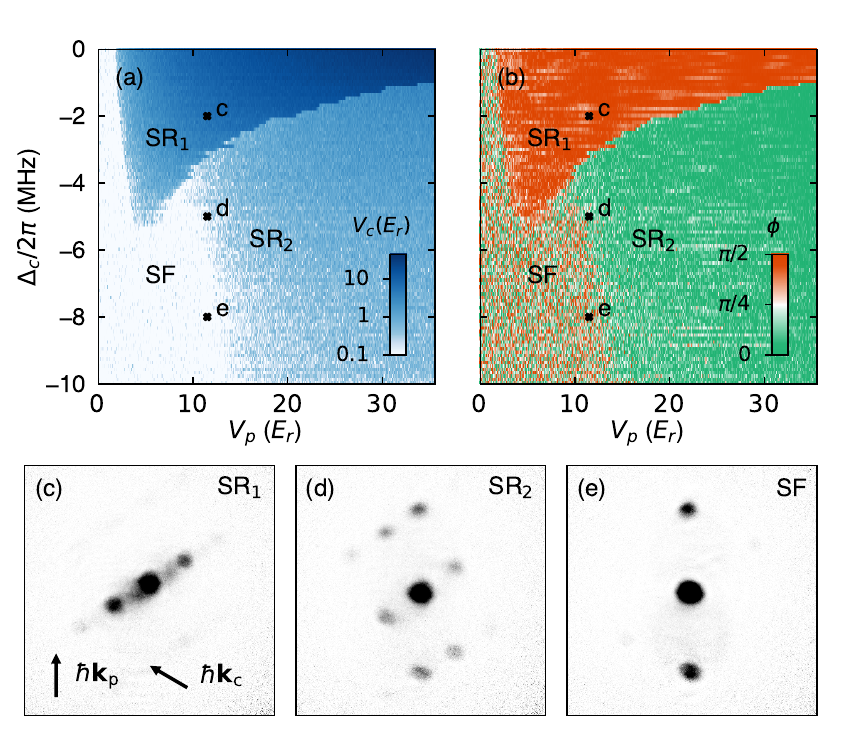}
\caption{(a, b) Phase diagram recorded with a heterodyne detector. The complex order parameter is the expectation value of intracavity field $\braket{\hat a}= |\alpha| e^{i\phi}$, measured as a function of the pump lattice depth $V_p$ and the cavity detuning $\Delta_c$. From the field amplitude $|\alpha|$ we extract the intracavity lattice depth $\braket{\hat{V}_c}= V_c$, which is plotted in (a). At the phase transition between the two superradiant phases the lattice depth $V_c$ changes abruptly. (b) Phase of the light field mapped to the first quadrant ($\phi \in [0,\pi/2]$) to highlight the $\pi/2$ phase jump. (c, d, e) Absorption images of the atomic cloud after a ballistic expansion of 27 ms, showing the momentum distribution of the atoms. The images are recorded at $V_p = 12(1)\, E_r$ and $\Delta_c/2\pi = -2, -5, -8$ MHz, as indicated in the phase diagrams. Dark areas show high atomic densities. (c, d) correspond to the atomic density distribution of Fig.~\ref{fig1} (a, c), respectively. The arrows denote the pump and cavity wave vectors $\hbar \kp$ and $\hbar \kc$.}
\label{fig2}
\end{figure}

We prepare a BEC of $N=4.8(4)\times 10^5$ $^{87}$Rb atoms and couple it dispersively to a single mode of our optical cavity~\cite{SM}. The pump beams have a wavelength of $\lambda = 780.1$~nm, which is blue detuned with respect to the D$_2$ line of $^{87}$Rb by $+2\pi\times76.6(1)$~GHz such that the atoms experience a repulsive potential. At this wavelength, the single atom dispersive shift is $U_0= 2\pi\times 43.6$ Hz, and the recoil frequency with one photon is $\omega_r = E_r/\hbar= 2\pi\times 3.77$~kHz. We vary the pump to cavity detuning $\Delta_c/2\pi = (\omega_p - \omega_c)/2\pi$ in a range of $0$ to $-10$~MHz, where $\omega_p$ and $\omega_c$ are the frequencies of pump beam and bare cavity resonance respectively. The two counter-propagating pump beams are incident on the atoms at an angle of $60(1)^\circ$ with respect to the cavity mode.

To characterize the system, we record phase diagrams as a function of detuning $\Delta_c$ and pump lattice depth $V_p$ in the following way. We initially fix the relative coupling strengths of the two counter-propagating pump beams by choosing $\gamma = 1.25(4)$.
We linearly ramp up the pump beam lattice depth $V_p$ from $0$ to $36(3)\, E_r$ in 50 ms and repeat the same experimental sequence for different values of $\Delta_c$. We record the light field leaking out of the cavity via a heterodyne detection setup and extract the field amplitude $|\alpha|$ and the phase $\phi$ as functions of $V_p$ and $\Delta_c$.
The resulting phase diagrams (Fig.~\ref{fig2} (a)) show three different phases, the normal (superfluid) phase (SF) and the superradiant phases SR$_1$ and SR$_2$. In the SF phase, the heterodyne setup only detects the vacuum noise of the cavity mode. In phases SR$_1$ and SR$_2$, non-zero average intracavity fields are detected. While the phase SR$_2$ extends to large lattice depths $V_p$, the phase SR$_1$ has a finite extent for nonzero cavity detunings
due to the parity of the self-organization Hamiltonian with positive atomic detuning, where the cavity-atom coupling gets counteracted by the growing band gap of the pump lattice~\cite{zupancic2019pband}. Blue pump detuning giving a limited extent to SR$_1$ is necessary to make SR$_2$ possible.

\begin{figure}[b!]
\centering
\includegraphics[]{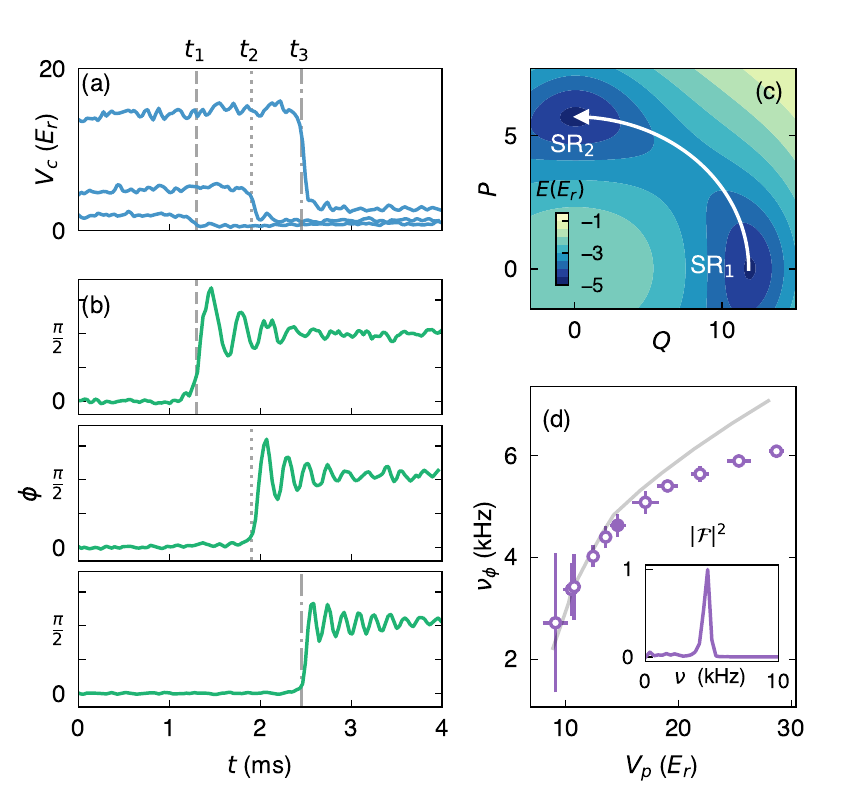}
\caption{Phase transition and relaxation of the order parameter. (a, b) Cuts of respectively the $V_c$ and the $\phi$ phase diagram (Fig.~\ref{fig2}~(a, b)) for different cavity detunings $\Delta_c/2\pi = - 3.75, - 2.75, - 1.75$ MHz, showing a jump at the phase transition. The transition points $t_{1,2,3}$ correspond respectively to $V_p= 11(1), 15(1), 25(2) E_r$. (c) The energy landscape truncated to the photonic space shows two minima on the different quadrature axes. The system jumps from the local minimum to the global minimum when the phase transition happens. We extract the curvature of the energy landscape at the global minimum (see supplementary information), which is plotted as the theoretical values in (d) (solid line). (d) We take the Fourier transform of the time trace of the oscillating phase, and extract the dominant frequency. The resulting frequencies are shown as a function of $V_p$ and accordingly changing $\Delta_c$, following the phase boundary between SR$_1$ and SR$_2$, and compared to the numerical model (grey solid line). The Fourier spectrum of the filled data point is shown in the inset. Errors indicate statistical deviation. }
\label{fig3}
\end{figure}

The $\pi/2$ difference in the phase of the cavity field between SR$_1$ and SR$_2$ (Fig.~\ref{fig2} (b)) is a consequence of the coupling to two orthogonal quadratures, corresponding to the two interference terms in Eq~\eqref{eq:HamiltonianAR} each representing one of the two crystal structures (See Fig.~\ref{fig1}). As is shown in Fig.~\ref{fig2}~(c, d, e), the difference also appears in time-of-flight images, where one records the momentum distribution of the atoms. In the normal phase, only the two momenta at $\pm 2 \hbar \k_p$, associated with the $\lambda/2$ periodicity of the pump lattice, are visible besides the zero-momentum mode, see Fig.~\ref{fig2} (e).
In SR$_1$, these momentum components are suppressed but the momenta $\pm \hbar (\k_p-\k_c)$ are populated, indicating a dominantly 1D density modulation, see Fig.~\ref{fig2} (c). In SR$_2$, two additional non-parallel momenta $\pm \hbar (\k_p+\k_c)$ are macroscopically populated, which results in an emergent 2D modulation.

In the transition from SR$_1$ to SR$_2$, the discontinuity of the order parameter is a first indication of a first-order phase transition. We plot the amplitude and phase of the cavity field as functions of time in Fig.~\ref{fig3} (a,~b) for different values of $\Delta_c$. In addition to the abrupt change of both observables at the phase transition, we record an oscillation of the phase $\phi$ after the transition. It has a single frequency that depends on $\Delta_c$ and $V_p$, and decays within a few oscillation periods, see Fig.~\ref{fig3} (b,~d).

Our observations can be understood as a transition from a metastable state to the ground state. We numerically calculate the energy landscape of the system from the Hamiltonian~\cite{SM}. It is plotted in Fig.~\ref{fig3} (c) as a function of the cavity field quadratures. There are two different minima located at the quadrature axes, corresponding to the two structural phases. Small changes in system parameters can turn the local minimum into a global minimum and vice versa. After ramping a control parameter, the system thus can temporarily be in the local minimum, but will eventually jump to the global minimum if the energy barrier is small. The oscillation in the phase of the light field after the transition reveals a collective excitation in SR$_2$. The excitation of this phase mode originates from the energy difference between the minima for SR$_1$ and SR$_2$ in the energy landscape when the transition takes place. We compare the oscillation frequencies with the expected frequencies of the phase mode in SR$_2$ calculated from the curvature of the energy landscape, which shows good quantitative agreement for low pump lattice depths, before interactions and atom loss lead to discrepancies at higher lattice depths (see Fig.~\ref{fig3} (d)).

\begin{figure}[h!]
\centering
\includegraphics[]{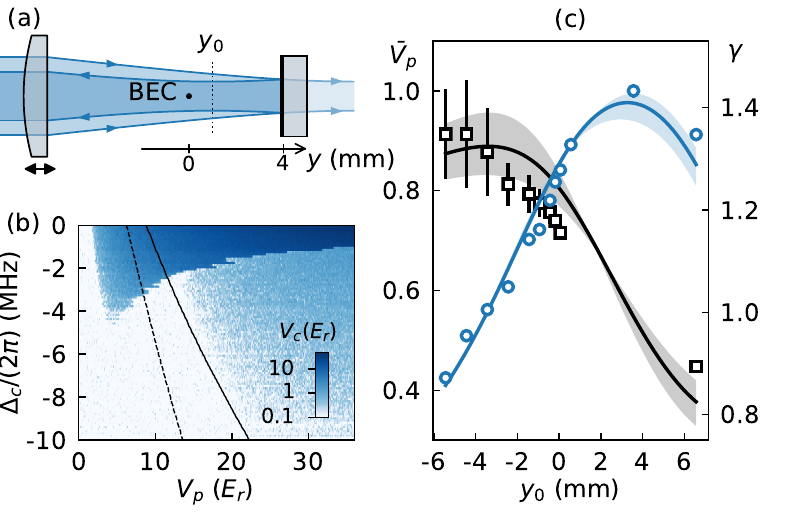}
\caption{Tuning the imbalance parameter $\gamma$. (a) Geometry of the system. The BEC is placed at a distance of around 4 mm from the retro-reflecting mirror. We choose the BEC position as the origin of the $y$ axis, and define $y_0$ as the coordinate of the beam focus. Moving the lens in $y$-direction translates the focus and allows tuning of $\gamma$. The mirror has a finite reflectivity, which corresponds to a minor shift in $y$ of the balanced point $\gamma =1$. (b) We extract $\gamma$ from the experimental data by comparing the threshold with numerical calculations. This phase diagram is consistent with an imbalance parameter $\gamma = 1.20$, which gives the calculated phase boundary of SR$_2$ shown as the solid curve. The dashed line is instead the calculated phase boundary for $\gamma = 1.25$ as in Fig.~\ref{fig2}. (c) Dependence of the pump lattice depth (blue curve) and imbalance parameter $\gamma$ (black curve) on the beam focus coordinate, calculated from the geometrical model of a retro-reflected Gaussian beam. The normalized pump lattice depth $\bar{V}_p$ is relative to the maximum measured value of $V_p$. The shaded areas account for the 5 \% systematic error of the beam waist measurement. The normalized pump lattice depths (blue circles) are measured by performing Raman-Nath diffraction calibration for different $y_0$ values. The imbalance parameters $\gamma$ (black squares) are extracted as in (b). Error bars account for the standard errors of the least square fits, and are smaller than the symbol size for the blue data.}
\label{fig4}
\end{figure}

In order to explore the parameter space of our theoretical model in Eq~\eqref{eq:HamiltonianAR} and Eq~\eqref{eq:V}, we record phase diagrams with different values of the imbalance parameter $\gamma$. Figure~\ref{fig4}~(a) shows how we experimentally tune $\gamma$: the initially collimated incident pump beam $\mathbf{E}_+$ is focused by a lens which generates a gaussian beam with a beam waist of \SI{38(2)}{\micro\metre}. The beam passes through the atomic cloud and is then retro-reflected again onto the atoms as $\mathbf{E}_-$ by an in-vacuo mirror, which itself is placed at a distance of about $4$~mm from the BEC, so the total travel of the pump beam is large compared to its Rayleigh range ($4.9$~mm). The relative coupling strength to the two pump fields can thus be continuously tuned by moving the lens which shifts the focus position of the pump beam.
The expected value of $\gamma$ can be calculated with gaussian optics, as we show in Fig.~\ref{fig4}~(c). From the beam geometry, we also calculate the lattice depth $V_p$ as a function of the lens position and compare it with Raman-Nath diffraction measurements. The highest lattice depth occurs as expected where the laser beam has its focus on the mirror.
The imbalance parameter $\gamma$ is obtained experimentally from the self-organization critical coupling: the phase boundary of the SF to SR$_2$ transition shifts to higher $V_p$ values by increasing the imbalance. We extract $\gamma$ by fitting the phase boundary (Fig.~\ref{fig4} (b)) with the numerical result from our theoretical model (\cite{SM}). The calculated and the fitted values of $\gamma$ are displayed in Fig.~\ref{fig4} (c), quantitative agreement eventually is limited by beam imperfections.

In conclusion, we explored a first order phase transition between two configurations of a self-organized superradiant crystal coupling to a single mode of an optical cavity. The real-time access to the intra-cavity field allowed us to study the relaxation behavior of this non-adiabatic structural phase transition. Our work demonstrates that quantum simulations with ultracold atoms not only provide conceptual insights into the electronic properties of a material \cite{Bloch2008Many, Lewenstein2012}, but can also be used to study lattice distortions and structural phase transitions. A natural extension of this method, complementary to the use of multi-mode resonators, is to use multiple pump beam modes in order to realize complex crystalline structures.

\begin{acknowledgments}
We gratefully acknowledge the support provided by Alexander Frank with the electronics and a valuable discussion with Nicola Spaldin. We acknowledge funding from SNF: project numbers 182650 and 175329 (NAQUAS QuantERA) and NCCR QSIT, from EU Horizon2020: ERCadvanced grant TransQ (project Number 742579) and ITN grant ColOpt (project number 721465), from SBFI (QUIC, contract No. 15.0019),  from EPSRC grants EP/P009565/1 and EP/K030094/1, and by an Investigator award of the Simons Foundation.

X.L. and D.D. contributed equally to this work.

\end{acknowledgments}

\beginsupplement
\section*{Supplementary material}

\subsection*{Experimental details and calibrations}

We prepare a cloud of $N=4.8(4)\times 10^5$ $^{87}$Rb atoms in the hyperfine state $\ket {F=1, m_F=-1}$. The atoms are held in a far off-resonant optical dipole trap with trapping frequencies $[\omega_x, \omega_y, \omega_z] = 2\pi\times$[104(1), 61(3), 195(5)] Hz. The optical cavity has a decay rate of $2\pi\times147(4)$ kHz. More details on the cavity and the locking scheme can be found in reference~\cite{zupancic2019pband}.
An offset field of $\sim25$~G is applied to avoid spin-dependent effects on self-organization due to the birefringence of our cavity~\cite{morales2019twomode}.

Both the pump lattice depth $V_p$ and the cavity lattice depth $V_c$ are calibrated with Raman-Nath diffraction. In the latter case, our heterodyne detection bandwidth is insufficient to measure pulses of few microseconds. We therefore calibrate the cavity lattice depths against a fast Single-Photon Counting Module (SPCM), and the SPCM versus the heterodyne at longer timescales.

As presented in the main text, tuning the imbalance parameter $\gamma$ is achieved by mechanically shifting the position of the focussing lens along the optical axis of the pump beam via a motorized stage (Physik Instrumente Q-522 and E-873). At every new position of the focusing lens, we additionally scan the lens position in the transverse directions and collect self-organization phase diagrams to ensure the maximum coupling of the pump beam to the cloud.

\subsection*{Phase resolved detection of the cavity light field}
A balanced optical heterodyne detection~\cite{carleton1968balanced} is performed to record not only the amplitude of the intracavity field but also the phase with respect to the transverse pump. The heterodyne regime is used instead of homodyne to avoid flicker noise in the electronics~\cite{yuen1983noise}, and the balanced detection is used to subtract excess laser intensity noise~\cite{stierlin1986excess}.

\begin{figure}[t!]
\centering
\includegraphics[]{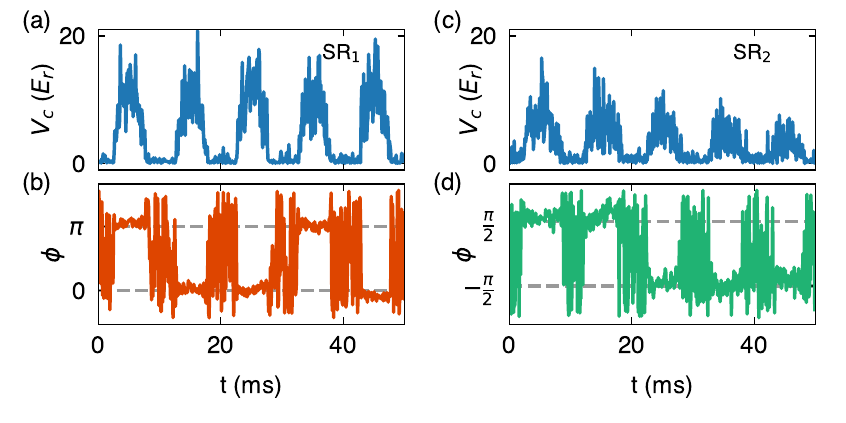}
\caption{$\mathbb{Z}_2$ symmetry breaking in the two superradiant phase transitions. The pump lattice depth $V_p$ is ramped up and down across the phase transition 5 times in a row during 50 ms and the phase $\phi$ of the intracavity field is recorded. Every time the critical point is crossed, the average intracavity field takes a finite value (a, c) and its phase with respect to the pump beam locks to either $(0,\pi)$ or $(\pi/2,-\pi/2)$ depending on whether the SR$_1$ (b) or SR$_2$ (d) transition point is crossed. Repeating the experiment multiple times shows that the selection of the phase is random~\cite{Baumann2011}. Data are taken at $\Delta_c = 0$ MHz (SR$_1$) and $\Delta_c = -2\pi\times6$ MHz (SR$_2$). }
\label{figsup1}
\end{figure}

To perform the heterodyne detection, a local oscillator light beam (LO), which is $70\,\text{MHz}$ lower in frequency than the transverse pump light, is combined on a beam splitter with the light field leaking from the cavity mirror and guided together onto the two photodiodes of a balanced photodetector (Thorlab PDB435A). The LO light is generated by shifting the frequency of the same laser source as the transverse pump light with an acousto-optical modulator and afterwards phase-locking it to the transverse pump. The LO optical power is regulated, resulting in \SI{30}{\micro\watt} power on each photodiode, which is sufficient for a shot noise limited detection~\cite{holmes1995optimum}. The output signal ($70\,\text{MHz}$ beating signal) of the balanced photodetector is then amplified, frequency converted down to $50\,\text{kHz}$, low-pass filtered and recorded by a PC oscilloscope (PicoScope 5444B) with a sampling rate of $1\,\text{MS/s}$. The recorded $50\,\text{kHz}$ signal data is processed by a computer program which extracts the two quadrature components of the signal via a digital IQ mixer. We rewrite the quadratures as amplitude and phase of the signal and then low-pass filter it by choosing a binning window of $1\times10^{-4}$ s. All the related radio frequency signal sources are phase locked to a $10\,\text{MHz}$ GPS frequency standard which has a fractional stability better than $10^{-12}$ to minimize the technical phase noise in the heterodyne detection system.

The measured absolute value of the phase $\phi$ of the intracavity field is determined by technical fluctuations in-between experimental repetitions. Nevertheless, the value relative to the pump field is fixed: when the system enters the superradiant phase SR$_1$ (SR$_2$), the phase $\phi$ is locked to either 0 or $\pi$ ($\pi/2$ or $-\pi/2)$, as it can be seen by recording multiple phase transitions in the same experimental run (see Fig.~\ref{figsup1}). However, since the lines in the phase diagrams consist of independent measurements, we plot $\phi$ modulo $\pi$ in Fig.~\ref{fig2}, \ref{fig3} and \ref{figsup2}. To get an absolute value for the phase $\phi$ in Fig.~\ref{fig2}~(b), we reference every line (i.e. every detuning $\Delta_c$) of the phase diagram to the measured $\phi$ at the highest $V_p$.

\subsection*{Interaction Hamiltonian and numerical results}

\begin{figure}[b!]
\centering
\includegraphics[]{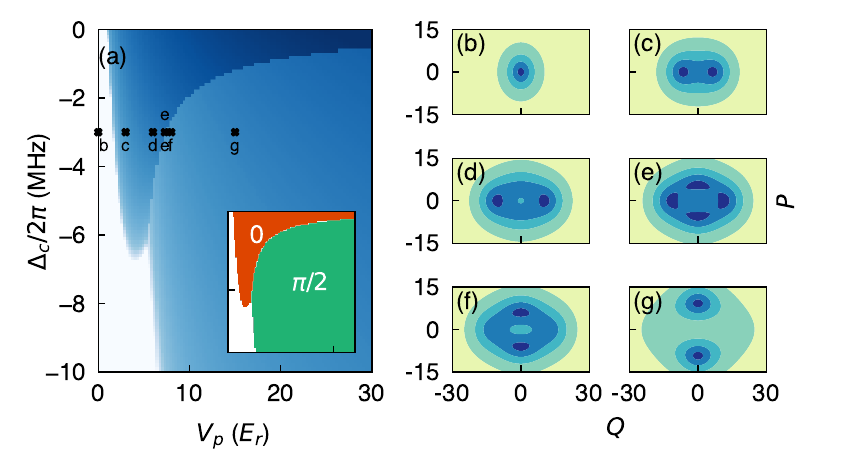}
\caption{(a) We plot the numerical calculated intracavity field as a function of the pump lattice depth $V_p$ and the cavity detuning $\Delta_c$ for the imbalance parameter $\gamma=1.4$. The inset shows the phase of the light field in the same parameter range. The points (b-g) correspond to the free energies plotted on the right as a function of the cavity field quadratures $Q$ and $P$.  From (b) to (c, d) the system enters SR$_1$ and two distinct minima appear on the real axes. The system spontaneously breaks this symmetry by selecting one of the two minima (see Fig.~\ref{figsup1}) Further increasing $V_p$, two new minima appear on the imaginary axis (e), corresponding to SR$_2$, which eventually become the global minima at even higher pump powers (f, g).}
\label{figsup2}
\end{figure}

The interaction between the atoms and the light is described by the atom-light Hamiltonian $\hat H_{A-L}  = -\alpha_s \hat{ \mathbf {E}}^*\cdot \hat{ \mathbf {E}}$, where $\alpha_s$ is the scalar atomic polarizability and $\hat{ \mathbf {E}}$ is the total electric field operator. The vectorial polarizability is zeroed by choosing the light field polarization to be parallel to the offset magnetic field along $\ez$. The field consists of two parts $\hat{ \mathbf {E}} = \hat{ \mathbf {E}}_c + \mathbf {E}_p$: the (quantum) cavity field $\hat{ \mathbf {E}}_c = \Ec \a \cos(\k_c\r) \ez$ and the (classical) transverse pump field,
which is given by two counterpropagating plane waves $\mathbf {E}_p = E_+ e^{i\k_p\r} \ez + E_- e^{-i\k_p\r} \ez$.
These fields lead to the following atom-light interaction Hamiltonian
\begin{equation}
\begin{split}
\hat H_{A-L}
&= -\alpha_{s} E_+E_- \cos^2(\k_p\r) - \alpha_{s} \Ec^2 \ad \a \cos^2(\k_c\r)  \\
& - \alpha_{s}E_0 (E_++E_-) (\a + \ad) \cos(\k_p\r)\cos(\k_c\r) \\
& +i  \alpha_{s} E_0(E_+-E_-) (\ad - \a) \sin(\k_p\r)\cos(\k_c\r),
 \end{split}
\label{eq:HSI}
\end{equation}
where we omitted a global energy shift. We introduce the pump and the cavity lattice potentials $\Vp = -\alpha_{s} E_+E_-$ and $V_c = \hbar U_0 \ad\a= -\alpha_{s} \Ec^2$, where $U_0$ is the dispersive shift of the cavity. For the atomic detuning used in this article $+2\pi\times76.6$ GHz and linearly polarized light, we have $U_0 = 2\pi\times 43.6$ Hz.
Defining $\gamma = (E_+ /E_-)^{1/2}$ one gets Hamiltonian~\eqref{eq:HamiltonianAR} with the potentials defined in~\eqref{eq:V}. At the phase transition, the system undergoes a change of the local symmetry, where the point of inversion symmetry for the real-space potential is shifted. Cavity decay is neglected, since for the parameters used in the present study it only leads to a minor shift $\propto \tan^{-1} (\kappa/\Delta_c)$ of the phase of the intra-cavity field.

To get a quantitative understanding of the phase transition, we numerically calculate the energy of the system as the mean-field expectation value of the Hamiltonian (neglecting interatomic interactions):
\begin{equation}
\hat H = -\hbar \Delta_{c} \hat a^\dagger \hat a + \frac{\hat{p}^2}{2m} + \hat H_{A-L},
\end{equation}
where $\hat{p}$ is the momentum operator and $m$ is the mass of a Rubidium atom. The atomic wave function is decomposed in the basis of momenta coupled to the BEC by the Hamiltonian, i.e. in second quantization formalism $\hat{\Psi}^\dagger = \sum_i \hat{c}_i^\dagger \psi_{\mathbf{k}_i}$, with the operator $\hat c^\dagger_i$ creating a particle in the $i-$th momentum state $\psi_{\mathbf{k}_i}$. We then consider the mean-field limit of Hamiltonian~\eqref{eq:HSI}, taking the expectation values for the operators and discarding higher order correlations, and numerically minimize the obtained mean field energy using $\braket{\hat c^\dagger_i} = \phi_i$ and $\braket \a = \alpha$ as variational parameters. Repeating this procedure for different cavity detunings $\Delta_c$ and pump lattice depths $V_p$, we get the numerical phase diagram of Fig.~\ref{figsup2}. The values of $\gamma$ in Fig.~\ref{fig4}(c) are obtained by optimizing $\gamma$ in the numerics to have best agreement between calculated and measured phase boundaries. In addition, we calculate and diagonalize the Hessian matrix of the energy at the minima and extract the elementary excitation energies of the system as the local curvature, $\omega^2 \propto \partial^2\mathcal{H}/(\partial {\alpha, \phi_i})^2$. The proportionality constant is given by the condition that $\omega(V_p=0,\alpha=0)$ is given by the energies of the bare momentum modes. The resulting excitation energies are shown in Fig.~\ref{fig3} of the main text.

\subsection*{Geometrical model for tuning the imbalance parameter}

Based on a geometrical model of a retro-reflected Gaussian beam, we calculate the pump lattice depths and the imbalance parameter $\gamma$ as functions of the beam focussing position, as is shown in Fig.~\ref{fig4}:

\begin{equation}
\begin{split}
V_p & = -\alpha_{s} \frac{\sqrt{R} E_0^{2} w_0^{2} }{w_+(y_0)w_-(y_0)} exp(-\frac{\delta r^{2}}{w_+(y_0)^{2}}-\frac{\delta r^{2}}{w_-(y_0)^{2}}) \\
\gamma & = \sqrt{\frac{w_-(y_0)}{R w_+(y_0)}} exp(-\frac{\delta r^{2}}{w_+(y_0)^{2}}+\frac{\delta r^{2}}{w_-(y_0)^{2}})\\
w_+(y_0) & = \sqrt{w_0^{2} + (\frac{\lambda}{\pi w_0})^{2} y_0^{2}}\\
w_-(y_0) & = \sqrt{w_0^{2} + (\frac{\lambda}{\pi w_0})^{2} (2 d - y_0)^{2}}\\
 \end{split}
 \label{eq:V}
\end{equation}

Here, $w_0$ and $E_0$ are the beam radius and electric field amplitude at the beam waist respectively. $y_0$ is the coordinate of the beam waist when choosing the BEC position as the origin. $R = 0.95$ is the reflectivity of the in-vacuo mirror at the wavelength of the pump laser. $\delta r$ is the possible residual radial distance from the BEC to the center axis of the pump beam. $w_+(y_0)$ and $w_-(y_0)$ are the radius of the incident and retro-reflected beam respectively at the BEC location as a function of the focus position. $d$ is the distance between the mirror and the BEC.

\FloatBarrier
\newpage
\bibliographystyle{apsrev4-1}

\end{document}